\documentclass[sigconf, nonacm]{acmart}

\usepackage{pvldb}% VLDB/PVLDB formatting overrides, see pvldb.sty for details.
\renewcommand\vldbdoi{}
\renewcommand\vldbpages{}
\renewcommand\vldbavailabilityurl{}
\renewcommand{\vldbtopmatter}{\pagestyle{\vldbpagestyle}}

\usepackage{xspace}
\usepackage{tabularx}
\usepackage{array}
\usepackage{enumitem}
\usepackage{placeins}
\usepackage{subcaption}

\usepackage{algorithm}
\usepackage{algpseudocode}

\graphicspath{{figures/}}
\setlist[itemize]{leftmargin=*,topsep=3pt,itemsep=2pt}
\newcommand{\sys}{\mbox{ZeroGR}\xspace}

\usepackage[framemethod=tikz]{mdframed}

\begin{document}
\title{\texorpdfstring{\sys}{ZeroGR}: Filtering-First Constrained Decoding for Generative Recommendation}

%%
%% The "author" command and its associated commands are used to define the authors and their affiliations.
\author{Xiangrui Yang}
\affiliation{%
  \institution{The University of Hong Kong}
  \city{Hong Kong}
  \country{China}
}
\email{xiangruiyang@connect.hku.hk}

\author{Jiawei Yang}
\affiliation{%
  \institution{Peking University}
  \city{Beijing}
  \country{China}
}
\email{yangjiawei26@stu.pku.edu.cn}

\author{Liang Zeng}
\affiliation{%
  \institution{Kuaishou Technology}
  \city{Beijing}
  \country{China}
}
\email{zengliang03@kuaishou.com}

\author{Daqian Ding}
\affiliation{%
  \institution{The University of Hong Kong}
  \city{Hong Kong}
  \country{China}
}
\email{daqianding@connect.hku.hk}

\author{Ziqian Liu}
\affiliation{%
  \institution{The University of Hong Kong}
  \city{Hong Kong}
  \country{China}
}
\email{liuziqian@connect.hku.hk}

\author{Bing Tian}
\affiliation{%
  \institution{Huazhong University of Science and Technology}
  \city{Wuhan}
  \country{China}
}
\email{tbing@hust.edu.cn}

\author{Cheng Peng}
\affiliation{%
  \institution{Kuaishou Technology}
  \city{Beijing}
  \country{China}
}
\email{pengcheng09@kuaishou.com}

\author{Yunfeng Zhao}
\affiliation{%
  \institution{Kuaishou Technology}
  \city{Beijing}
  \country{China}
}
\email{zhaoyunfeng@kuaishou.com}

\author{Fangyu Zhang}
\affiliation{%
  \institution{Kuaishou Technology}
  \city{Beijing}
  \country{China}
}
\email{zhangfangyu@kuaishou.com}

\author{Jiaqiang Liu}
\affiliation{%
  \institution{Kuaishou Technology}
  \city{Beijing}
  \country{China}
}
\email{liujiaqiang@kuaishou.com}

\author{Xiao Liang}
\affiliation{%
  \institution{Kuaishou Technology}
  \city{Beijing}
  \country{China}
}
\email{liangxiao03@kuaishou.com}

\author{Yiming Qiu}
\affiliation{%
  \institution{The University of Hong Kong}
  \city{Hong Kong}
  \country{China}
}
\email{yimingq@hku.hk}

\begin{abstract}
Generative recommendation is gaining traction in production, retrieving items by generating their identifiers one token at a time. The generated sequences must identify relevant catalog items, so constrained decoding couples relevance scoring with validity filtering. Because most possible sequences are invalid, scoring before filtering wastes computation on candidates that will be discarded. Our evaluation shows that filtering first can avoid this work with negligible impact on recommendation quality. We present \sys, a filtering-first constrained decoding system for generative recommendation. \sys uses a \textit{step-sliced filtering index} to reduce GPU memory use by tailoring data representations to each decoding step. Its execution engine adopts \textit{shared-slot execution} to accelerate data exchange across stages and \textit{hybrid projection} to adapt computation to workload patterns, thereby reducing execution overhead. During index refreshes, \sys applies \textit{layout-guided streaming construction} to speed up updates and limit serving interference. Extensive evaluation shows that \sys accelerates the filtering and scoring pipeline by up to $9.9\times$ over representative baselines while reducing GPU index memory use and accelerating index reconstruction.
\end{abstract}

\maketitle

\vldbtopmatter

\section{Introduction}
\label{sec:introduction}

Recommendation systems (RecSys) underpin online services, selecting relevant items for users from enormous catalogs under interactive latency constraints. Inspired by the rapid advancement of LLMs, emerging generative recommendation (GR) approaches item discovery by autoregressively generating item identifiers from user context~\citep{decao2021genre,tay2022dsi}. These GR models promise better recommendation quality and less operational toil than conventional RecSys, and have gained traction in production deployments~\citep{deng2025onerec,chen2026onesearch,onerec2026onereason}. To cover catalogs with millions of items, GR represents each item as a short sequence of semantic identifier (SID) tokens, with related items sharing sequence prefixes~\citep{rajput2023tiger,singh2024semanticids}. Decoding these tokens one at a time progressively narrows the item space toward specific recommendations. Because each step depends on the preceding tokens, its cost adds directly to the user request latency~\citep{lin2025atspeed}.

To find multiple relevant items, GR uses beam search to explore competing partial SID sequences, \emph{scoring} their continuations and retaining promising candidates at each decoding step~\citep{rajput2023tiger,deng2025onerec}. These scores guide \emph{relevance} to the user, but cannot establish \emph{validity} within the catalog: a seemingly promising SID sequence may identify no actual item. Each GR decoding step therefore also requires \emph{filtering}, typically through a catalog trie that restricts continuations to those that can reach valid items~\citep{zheng2024lcrec,li2025leadre,su2026static}. This search for relevant recommendations subject to catalog validity is the constrained decoding problem in GR. Maintaining competing item candidates requires both scoring and filtering to process many continuations at every step, and their decisions determine which candidates advance. Both operations thus lie on the serving critical path, making their performance central to meeting GR's latency objective.

The order of scoring and filtering directly affects how much work GR decoding performs. \textit{Scoring first}~\cite{su2026static} scores all possible continuations before discarding invalid ones, whereas \textit{filtering first}~\cite{dong2025xgrammar} restricts scoring to valid continuations permitted by the catalog. At first glance, filtering first is attractive: valid candidate sparsity allows it to eliminate substantial scoring work while preserving the same validity guarantee. Its effect on relevance, however, is less straightforward. 
Scoring first normalizes scores over all continuations of each sequence prefix, whereas filtering first normalizes only over valid continuations. Depending on the total probability assigned to invalid continuations under each prefix, filtering first can give candidates from some prefixes a larger boost than others, changing their relative ranking and ultimately which candidates survive beam search.
The two orders can therefore return different recommendations even though both enforce the same catalog constraints. 
The key question therefore becomes \textit{whether filtering first can deliver recommendation quality on par with scoring first}.

In this paper, we answer this question in the affirmative. Our evaluation shows virtually identical recall for the two orders, with approximately $99.9\%$ overlap in the returned items. This stability is consistent with how GR progressively narrows the search: early steps assign most probability mass to valid continuations, so filtering has little effect on normalization; later steps mainly refine choices established earlier, so changes in normalization tend to have limited impact on recommendation quality. We therefore aim to build a filtering-first constrained decoding pipeline for GR serving. The remaining question is \textit{how to realize the performance benefits of filtering first} given the following coupled challenges:

\begin{itemize}
    \item \textit{Representing filtering index within limited GPU memory.} 
    Filtering requires efficient access to valid continuations; keeping this lookup on the GPU reduces CPU--GPU coordination. The GPU filtering index must cover the catalog and accommodate varying valid candidate densities across decoding steps, while sharing memory with model weights and inference state~\citep{kwon2023pagedattention}. The challenge is to keep the index compact across diverse step profiles.

    \item \textit{Translating filtering-first computation into efficient GPU execution.} Computing fewer scores does not by itself ensure efficient GPU execution~\citep{xia2023flashllm}. Filtering first produces selective workloads that vary across prefixes and decoding steps, and the overhead of organizing and processing them can outweigh the scoring savings. The challenge is to adapt GPU execution to these workload patterns so that reduced scoring translates into lower latency.
    
    \item \textit{Building filtering index without disrupting GPU serving.}
    Because the catalog and SID assignments are refreshed with the model, each GR update requires rebuilding the filtering index. GPU construction competes with ongoing inference for compute, while transferring a CPU-built index consumes bandwidth also needed by serving. The challenge is to complete the rebuild within the update window while protecting the latency of ongoing requests.
\end{itemize}

We present \sys, a filtering-first constrained decoding system for generative recommendation. To save GPU memory, \sys uses a \textit{step-sliced filtering index}: for each decoding step, it maintains a mapping from sequence prefixes to valid next tokens, encoded in either a sparse or a dense representation according to that step's valid candidate density. Building on this index, the \sys execution model adopts \textit{shared-slot execution} to give each prefix fixed candidate slots shared by filtering, scoring, and candidate selection, allowing stages to exchange results directly and reducing coordination overhead. The execution model further applies \textit{hybrid projection}: when the scoring reduction outweighs selective execution overhead, it computes scores only for valid continuations; otherwise it computes all scores to preserve regular batched execution.
To speed up refreshes and limit serving interference, \sys applies \textit{layout-guided streaming construction}: it precomputes each fragment's size and position in the final index, allowing CPU workers to generate fragments independently in their final form while streaming completed fragments to the GPU.

We evaluate \sys on a variety of recommendation datasets. Compared with representative constrained decoding baselines, \sys reduces the combined latency of filtering, scoring, and candidate selection by up to $9.9\times$. It also reduces GPU index memory use by up to $74.5\%$ and reconstructs its index $278\times$ faster, completing the rebuild within the service's update window. In summary, this work makes the following contributions:

\begin{itemize}
  \item We analyze valid candidate sparsity in GR decoding and empirically demonstrate that filtering first can reduce scoring work while retaining comparable recommendation quality.
  \item We design \sys, integrating a compact filtering index, coordinated GPU execution, and parallel streaming construction to realize the performance benefits of filtering first.
  \item We evaluate \sys on real-world datasets from industrial GR services, demonstrating significant improvement over representative constrained decoding baselines.
\end{itemize}

\section{Background and Motivation}
\label{sec:background}

\subsection{Sparsity of Valid Sequences in Generative Recommendation}

\begin{figure}[t]
  \centering
  \includegraphics[width=\linewidth]{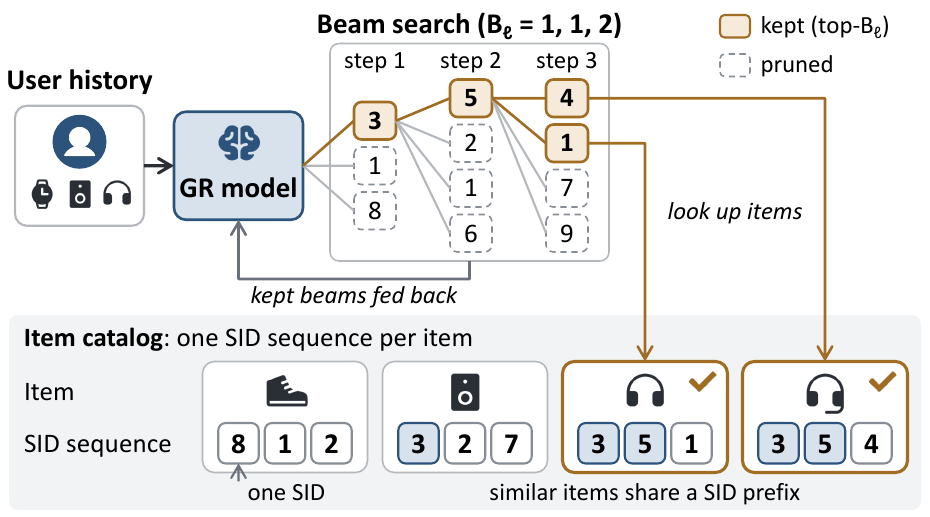}
  \vspace{-4mm}
  \caption{Generative recommendation with SID sequences.}
  \Description{A user's interaction history feeds a GR model. A beam-search panel shows three decoding steps with beam width two: step one keeps SIDs 3 and 8 and prunes 1; step two keeps 3-5 and 3-2 and prunes 8-1 and 8-6; step three keeps 3-5-4 and 3-5-1 and prunes 3-2-7 and 3-2-9. The kept beams are fed back to the model. The two completed sequences point to the headset (3, 5, 4) and the headphones (3, 5, 1) in an item catalog that also lists a sneaker (8, 1, 2) and a speaker (3, 2, 7); SIDs shared across items are highlighted.}
  \label{fig:gr-inference}
  \vspace{-6mm}
\end{figure}

The conventional paradigm in recommender systems decomposes item discovery into a multi-stage ``retrieve-and-rank'' pipeline~\citep{covington2016youtube}: first filtering a coarse candidate set from the catalog, and then applying increasingly complex models to rank them.
In contrast, generative recommendation (GR) departs from this pipeline by using an autoregressive model to directly generate the identifiers of target items, fundamentally casting item retrieval as sequence generation~\cite{rajput2023tiger}.
A key design choice in GR is how to represent item identifiers. 
Early GR systems assign each item a dedicated single token~\cite{zhai2024hstu}, which suffers from vocabulary explosion and fails to capture semantic similarity across items.
In contrast, modern SID-based GR~\cite{onerec2026onereason} represents each item $i$ with an $L$-token SID sequence $\mathbf{s}(i)=(s_1^{(i)},\ldots,s_L^{(i)})$, where each semantic identifier (SID) $s_\ell^{(i)}$ is a token drawn from a position-specific codebook $\mathcal{V}_\ell$. These sequences are constructed offline such that items with similar content share common SID prefixes~\cite{rajput2023tiger,singh2024semanticids,lee2022rqvae}. For example, two similar products (e.g., headphones and a headset) might share the coarse prefix $(3, 5)$ and differ only in their final token (e.g., $(3, 5, 1)$ and $(3, 5, 4)$ in Figure~\ref{fig:gr-inference}, bottom). 
As a result, each SID prefix represents a category of related items rather than a single item, and only a complete $L$-token sequence uniquely identifies a specific item in the catalog.

At serving time, item retrieval turns into an autoregressive decoding workload over these codebooks.
As shown at the top of Figure~\ref{fig:gr-inference}, the GR model first encodes the user's interaction history and then, starting from a beginning-of-sequence token, generates one SID per autoregressive decoding step, conditioning each prediction on the encoded context and the previously generated prefix.
To return multiple recommendations, decoding uses beam search~\cite{rajput2023tiger,deng2025onerec}, which maintains $B_\ell$ active prefixes, or beams, at each decoding step $\ell$, where the beam width $B_\ell$ can vary across steps.
After $L$ steps, each completed SID sequence is looked up in the item catalog $\mathcal{I}$ and resolved to its item.
Intuitively, the set of possible SID sequences is the Cartesian product of the per-step codebooks, $\mathcal{V}_1 \times \cdots \times \mathcal{V}_L$, whose cardinality $\prod_{\ell=1}^L |\mathcal{V}_\ell|$ is astronomical, far exceeding $10^{16}$ in production systems.
However, the set of actual valid SID sequences, $\mathcal{S} = \{\mathbf{s}(i) \mid i \in \mathcal{I}\} \subseteq \mathcal{V}_1 \times \cdots \times \mathcal{V}_L$, is only around $10^7$--$10^8$, meaning that the majority of possible sequences are invalid, i.e., they do not map to any items.
Navigating this vast, sparsely populated sequence space to find high-quality, valid items is the core task of GR decoding.

\subsection{Scoring and Filtering in SID Decoding}

To return high-quality recommendations that correspond to actual catalog items, each decoding step in GR requires both \emph{scoring for quality} and \emph{filtering for validity}.
For \emph{scoring}, at step $\ell$, the model's output projection computes logits over the full codebook $\mathcal{V}_\ell$ for each of the $B_\ell$ active beams, $B_\ell|\mathcal{V}_\ell|$ logits in total.
Beam search normalizes these logits into scores via softmax, adds the scores to the accumulated prefix scores, and retains the $B_{\ell+1}$ highest-scoring extensions for the next step.
For \emph{filtering}, decoding must ensure that generated sequences stay within the valid set $\mathcal{S}$, since relying on model scores alone can easily produce sequences that map to no item.
Deferring filtering until all $L$ steps finish, i.e., post-generation filtering~\cite{yang2026adabool}, wastes computation on invalid paths and requires heavy over-generation to fill the requested output size.
Instead, constrained decoding enforces validity at every step $\ell$ by restricting each active prefix $p = (s_1, \ldots, s_{\ell-1})$ to its valid SID continuations:
\begin{equation}
\operatorname{Next}_{\ell}(p) = \{v \in \mathcal{V}_\ell \mid \exists\, \mathbf{s} \in \mathcal{S},\; \mathbf{s}_{1:\ell-1} = p \;\wedge\; s_\ell = v\}.
\label{eq:valid-next-tokens}
\end{equation}
For instance, for prefix $p = (3, 5)$ in Figure~\ref{fig:gr-inference}, only $\operatorname{Next}_{3}(p) = \{1, 4\}$ can extend into valid items $(3, 5, 1)$ and $(3, 5, 4)$.
Because $\operatorname{Next}_{\ell}(p)$ varies across prefixes, systems build a \emph{filtering index} over $\mathcal{S}$ to retrieve $\operatorname{Next}_{\ell}(p)$ during decoding.

\begin{figure}[t]
  \centering
  \includegraphics[width=\linewidth]{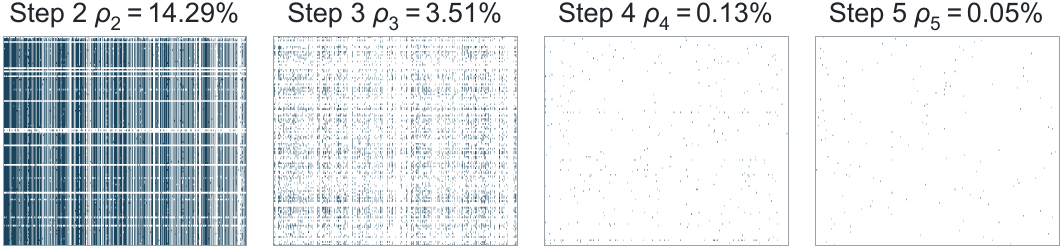}
  \vspace{-2mm}
  \caption{Valid SID candidates per decoding step.}
  \Description{Four matrices for decoding steps two through five, with valid candidate densities of 14.29, 3.51, 0.13, and 0.05 percent. Rows are active beam prefixes, columns are tokens, and blue cells mark valid SID candidates.}
  \label{fig:search-beam-state-sid}
  \vspace{-2mm}
\end{figure}

At each decoding step, filtering and scoring can be combined in two execution orders: \emph{filtering-then-scoring} removes the tokens outside $\operatorname{Next}_{\ell}(p_b)$ before normalizing logits into scores, so each beam is normalized \emph{only} over its valid tokens and the logits of the other tokens need not be computed at all, whereas \emph{scoring-then-filtering} first computes and normalizes all $B_\ell|\mathcal{V}_\ell|$ logits over the entire codebook $\mathcal{V}_\ell$ and then masks out the tokens outside $\operatorname{Next}_{\ell}(p_b)$.
For brevity, we call them \emph{filtering first} and \emph{scoring first}.
For performance reasons, filtering first is preferable because valid SID continuations are sparse.
We quantify the fraction of useful full-projection logits at step $\ell$ using the \emph{valid candidate density}:
\begin{equation}
  \rho_\ell = \frac{\sum_{b=1}^{B_\ell} |\operatorname{Next}_{\ell}(p_b)|}{B_\ell|\mathcal{V}_\ell|}.
  \label{eq:valid-candidate-density}
\end{equation}
The numerator counts the valid continuations summed over all $B_\ell$ active beams, whereas the denominator counts the $B_\ell|\mathcal{V}_\ell|$ logits that full projection computes, one per beam and token.
Hence, $\rho_\ell$ is the average fraction of the codebook that can still extend an active prefix, ranging from $1/|\mathcal{V}_\ell|$ when every prefix admits a single continuation to $1$ when every token remains valid.
As measured on our industrial generative search dataset in Figure~\ref{fig:search-beam-state-sid}, valid entries, shown in colored cells, account for at most $\rho_2 = 14.29\%$ at step 2 and drop to $\rho_5 = 0.05\%$ by step 5.
Thus, scoring first wastes $85.7\%$ to $99.95\%$ of its projection arithmetic, logit materialization, and masking overhead on invalid continuations, whereas filtering first allows the decoder to skip this redundant work.

In practice, a well-trained model rarely generates invalid tokens, but it does not make scoring first any cheaper.
The fraction of logits wasted on invalid tokens, $1 - \rho_\ell$, is fixed by the item catalog rather than by model accuracy, and it varies widely across decoding steps and across prefixes within a step.

\subsection{Viability of Filtering First}

In single-sequence generation such as standard LLM decoding~\citep{zheng2024sglang}, adopting filtering first is straightforward because both orders produce the exact same result: all candidate tokens compete within the same prefix, so excluding invalid tokens before or after scoring does not alter the relative ranking among valid tokens.
In GR, however, filtering first and scoring first are \emph{not} mathematically equivalent.
Unlike single-sequence decoding, GR relies on beam search across $B_\ell > 1$ active prefixes and ranks candidate extensions globally by adding each token's current-step log-softmax score to its accumulated prefix score.
Under scoring first, each beam's softmax denominator sums over the full codebook $\mathcal{V}_\ell$, penalizing prefixes that assign high probability mass to invalid tokens.
Under filtering first, because invalid tokens are never scored, the softmax denominator sums only over the valid subset $\operatorname{Next}_{\ell}(p_b)$~\citep{park2024gad}.
Since $|\operatorname{Next}_{\ell}(p_b)|$ varies across active prefixes $p_b$, filtering first normalizes each beam over a token set of a different size, so the two orders can rank the same candidate extensions differently.
Because scoring first keeps the full-codebook normalization of unconstrained decoding, it appears to be the more faithful order in GR, and existing constrained decoding systems for GR~\cite{rajput2023tiger,su2026static} simply adopt it as the default despite its high computational cost.

\begin{figure}[t]
  \centering
  \includegraphics[width=\linewidth]{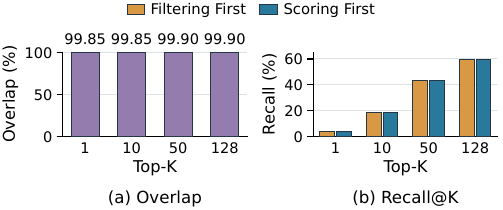}
  \vspace{-6mm}
  \caption{Overlap and recall of filtering first and scoring first on MovieLens-1M.}
  \Description{Two bar charts over top-K cutoffs of 1, 10, 50, and 128. Panel (a) shows that the lists returned by filtering first and scoring first overlap by 99.85, 99.85, 99.90, and 99.90 percent. Panel (b) shows Recall@K of the two orders as paired bars of nearly equal height, rising from about 4 percent at K of 1 to about 60 percent at K of 128.}
  \label{fig:order-overlap}
  \vspace{-6mm}
\end{figure}

The two orders differ sharply in computation but negligibly in recommendation quality.
Figure~\ref{fig:order-overlap} compares the two orders with a GR model trained on the public MovieLens-1M dataset~\cite{harper2015movielens}.
The items they return overlap by 99.90\% on average, and their recall is virtually identical at every cutoff.
This results from the step-wise sparsity pattern of SID decoding.
Across decoding steps, $\rho_\ell$ is relatively high in early steps and becomes extremely small only in the final steps, which is shown in Figure~\ref{fig:search-beam-state-sid}.
In the early steps, each active prefix still admits many valid continuations, and a trained GR model already concentrates nearly all of its probability mass on valid tokens; excluding invalid tokens from the softmax denominator therefore leaves the normalized step scores almost unchanged.
In the later steps, a candidate's total score comes mostly from the earlier steps, so the order of filtering and scoring at these steps rarely changes which candidates are kept.
As a result, GR recommendation quality is not sensitive to the ordering between filtering and scoring operations.

\subsection{Insights and Challenges}
\label{sec:insights}

\begin{figure*}[!t]
  \centering
  \includegraphics[width=0.92\linewidth]{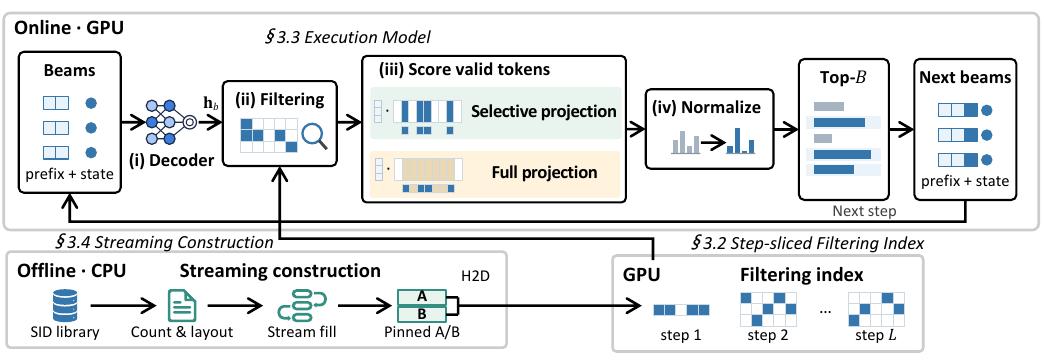}
  \vspace{-3mm}
  \caption{Overview of \sys: offline index construction and online filtering, scoring, and selection at each decoding step.}
  \Description{The item catalog is built offline into a filtering index, and each online decoding step filters the candidates of each beam, scores the valid tokens, and selects the next beams.}
  \label{fig:zerogr-overview}
  \vspace{-2mm}
\end{figure*}

The analysis above shows that the catalog fixes the computation space before any score is computed.
This fact leads to the following insights about how SID constraints can be used during decoding.

\textit{Insight 1: SID constraints specify computation rather than outputs.}
Because $\operatorname{Next}_{\ell}(p)$ is fixed by the item catalog at the offline construction phase, it can specify which computation the decoder performs.
Instead of masking scores after a full-codebook projection, the decoder retrieves $\operatorname{Next}_{\ell}(p_b)$ for each active beam first and computes logits only for these tokens.
A constraint used this way reduces the logits the projection computes from $B_\ell|\mathcal{V}_\ell|$ to $\rho_\ell B_\ell|\mathcal{V}_\ell|$ and removes constraint application as a post-processing stage.

\textit{Insight 2: the constraint is a family of per-step mappings.}
Because $\operatorname{Next}_{\ell}(p)$ is indexed by prefix, the naive representation is a prefix tree over $\mathcal{S}$.
Decoding, however, never queries the constraint as a whole: step~$\ell$ queries only prefixes of length $\ell-1$, each carried by an active beam from step~$\ell-1$.
For each such prefix $p$, the step needs only $\operatorname{Next}_{\ell}(p)$; appending the continuations that beam selection keeps yields the length-$\ell$ prefixes that step~$\ell+1$ queries.
The constraint is therefore a family of $L$ per-step mappings, and step~$\ell$ retrieves $\operatorname{Next}_{\ell}(p_b)$ for all $B_\ell$ active beams with one batched lookup into its own mapping.

The two insights determine what each decoding step needs to compute.
Realizing them within a GPU serving instance, however, raises three system-level challenges.

\textit{Challenge 1: supporting per-step heterogeneity under a GPU memory budget.}
Decoding steps favor different computations: when most tokens remain valid, scoring every token with one regular matrix product is efficient; when only a few remain, we should score just the valid candidates of each beam.
An index that serves every step should therefore provide candidates in the form each step's computation needs, and leave the choice of computation open rather than fixing it in advance.
At the same time, the index must stay small because it resides in GPU memory next to the model weights and the attention caches~\citep{zhang2025pqcache,chen2026retroinfer,wang2026codec}.
With more than $10^{8}$ items, every extra field per entry takes memory that would otherwise go to serving larger batches.
The challenge is to provide this per-step adaptability while keeping the index compact.

\textit{Challenge 2: executing on the index without forfeiting GPU efficiency.}
An index that identifies which entries to compute does not by itself make them cheap to compute, because the full projection it replaces is a regular batched matrix product while an index-driven projection reads a different, variable-length set of codebook entries for every beam.
Simply skipping the invalid entries can therefore leave latency unchanged or even increase it relative to full projection, despite performing much less arithmetic~\citep{gale2020sparse}.
The same irregularity carries over to beam selection.
The challenge is not only to reduce the computation but also to execute the remaining irregular work efficiently enough that the savings translate into lower latency.

\textit{Challenge 3: reconstructing the index alongside online serving.}
Because the item catalog and its SID sequences are refreshed together with the model weights, every model update must rebuild the index within a limited time window before the new version can serve.
The rebuild also runs while the instance continues to serve the current version.
The rebuild must therefore neither contend with inference for GPU compute, which inference requires to meet its latency targets, nor saturate the host-to-GPU link that carries incoming requests.
The challenge is to reconstruct the index within the update window while leaving both GPU compute and the host-to-GPU link to online serving.

\section{\texorpdfstring{\sys}{ZeroGR} Design}
\label{sec:design}

This section presents the design of \sys, beginning with an overview of its parts and request processing during decoding.

\subsection{Overview}
\label{sec:design-overview}

\sys filters the candidates of each beam through the index before scoring them, so that the item catalog, rather than the full codebook, determines which scores each decoding step computes.
Its design has three parts: a filtering index that lists the valid next tokens of each prefix (\S~\ref{sec:constraint-index}), an execution engine that uses the index at every decoding step (\S~\ref{sec:index-execution}), and a construction pipeline that builds this index (\S~\ref{sec:cpu-staged-construction}).

Figure~\ref{fig:zerogr-overview} gives an overview of \sys.
Each request expands into up to $B$ beams through beam search.
Each beam $b$ carries an active prefix $p_b$, an accumulated score $a_b$, and a state $q_b$, an integer that locates the valid next tokens of $p_b$ in the filtering index.
Each decoding step processes all beams of the request in four stages.
(i)~The decoder produces the hidden vector $h_{\ell,b}$ of each beam.
(ii)~\sys uses each beam's state to read its valid next tokens from $\mathcal{M}_\ell$, the part of the index for the current step $\ell$.
(iii)~\sys computes the logits of these valid tokens.
It does so in one of two modes: selective projection computes only their logits, whereas full projection, used when most tokens are valid, computes the logits of all tokens in $\mathcal{V}_\ell$ and keeps only the valid ones.
(iv)~Each beam normalizes its logits over its valid tokens and adds its accumulated score, and the $B$ highest-scoring candidates of the request become the beams of the next step.
After $L$ steps, each beam holds a complete SID sequence that identifies an item in the item catalog, and the request returns these $B$ items.

Whereas the online path reduces per-step decoding latency, index construction keeps the filtering index consistent with the item catalog, which is refreshed together with the model weights.
Each refresh must complete within a limited window while the instance continues to serve requests, so \sys rebuilds the index on the CPU and keeps the rebuild from competing with inference for GPU compute or host-to-GPU bandwidth.

\subsection{Index Representation}
\label{sec:constraint-index}

\begin{figure}[t]
  \centering
  \includegraphics[width=\linewidth]{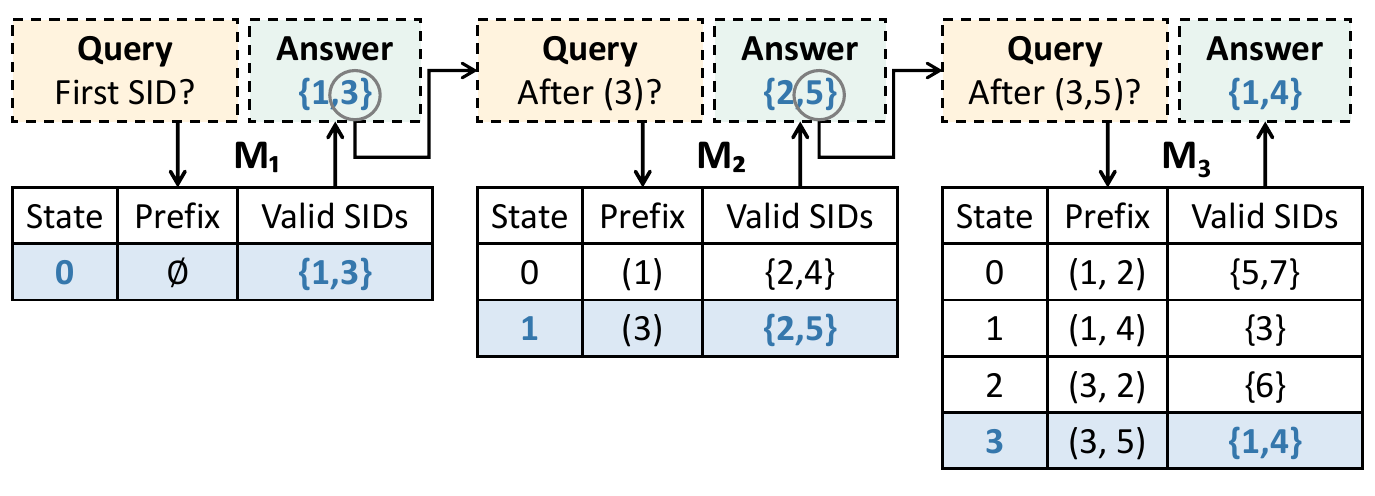}
  \caption{Step-sliced filtering index, where step~$\ell$ holds the prefixes of length $\ell-1$ and numbers them independently.}
  \Description{Six valid SID sequences are (1,2,5), (1,2,7), (1,4,3), (3,2,6), (3,5,1), and (3,5,4). Step one has state zero for the empty prefix. Step two has states zero and one for prefixes (1) and (3). Step three has states zero through three for prefixes (1,2), (1,4), (3,2), and (3,5). Arrows labeled 3 and 5 trace the empty prefix to step-two state one and then step-three state three.}
  \label{fig:step-sliced-index}
  \vspace{-4mm}
\end{figure}

\sys introduces a \emph{step-sliced filtering index}, which makes each decoding step independent in representation and lookup while keeping the index compact in GPU memory.
The index comprises $L$ mappings, $\mathcal{M}_1,\ldots,\mathcal{M}_L$, one per decoding step, each telling which tokens can follow a prefix at that step.
Slicing by step makes the mappings independent of one another, so different steps can use different representations.
For a prefix $p=(s_1,\ldots,s_{\ell-1})$, its state $q_\ell(p)$ identifies a row within $\mathcal{M}_\ell$.
For $\ell<L$, this row lists each valid next token of $p$ together with its successor state in $\mathcal{M}_{\ell+1}$:
\begin{equation}
  \mathcal{M}_{\ell}(q_{\ell}(p))
  = \{(v, q_{\ell+1}(p \circ v)) \mid v \in \operatorname{Next}_{\ell}(p)\},
  \label{eq:constraint-step-mapping}
\end{equation}
where $p\circ v$ appends token $v$ to $p$.
State identifiers are local to a step: the selected extension supplies the row to query in the next mapping, without traversing the preceding prefix again.
Figure~\ref{fig:step-sliced-index} shows the index for six items with three-token SID sequences.
Step~$1$ holds only the empty prefix, step~$2$ holds the prefixes $(1)$ and $(3)$, and step~$3$ holds the four two-token prefixes, with each step numbering its prefixes from zero.
For the items that begin with $(3,5)$, token~$3$ leads from step~$1$ to state~$1$ of step~$2$, and token~$5$ then leads to state~$3$ of step~$3$.
At step~$3$, this state alone gives the valid last tokens, $1$ and $4$, without revisiting tokens $3$ and $5$.
Each step thus provides either token-aligned validity information for full projection or a compact list of valid tokens for selective projection.

Specifically, a step with many valid continuations uses a \emph{dense mapping}, which provides token-aligned validity information through a successor table whose rows are states and whose columns are the SIDs of the step's codebook $\mathcal{V}_\ell$.
Each cell holds the successor state reached by appending the column's SID to the row's prefix, and a sentinel marks an invalid SID.
A cell need not store its SID, because its column already identifies it.
Since a row marks valid SIDs by position rather than listing them, a dense mapping serves full projection, whose logits are laid out in the same token order as the table columns.

A step with few valid continuations instead uses a \emph{sparse mapping}, which keeps only the valid SIDs as a compact list.
The list places the valid SIDs of all prefixes in one SID array $A_\ell$, one segment per state, with segments ordered by state.
A boundary array $R_\ell$ records where each segment begins and ends, so two adjacent boundaries delimit the valid SIDs of one state.
Invalid SIDs are skipped, while an SID that is valid under several prefixes appears once in each of their segments.
Each entry therefore stands for exactly one extended prefix, so a sparse mapping needs no successor table: the position of an entry serves as its successor state.
Since a segment lists its valid SIDs explicitly, a sparse mapping serves both projection modes: selective projection computes logits only for the listed SIDs, whereas full projection computes the logits of all tokens in $\mathcal{V}_\ell$ and uses the listed SIDs to pick out the valid ones.
For a sparse mapping, the execution engine chooses between the two modes without changing the representation.

\begin{figure}[t]
  \centering
  \includegraphics[width=0.9\linewidth]{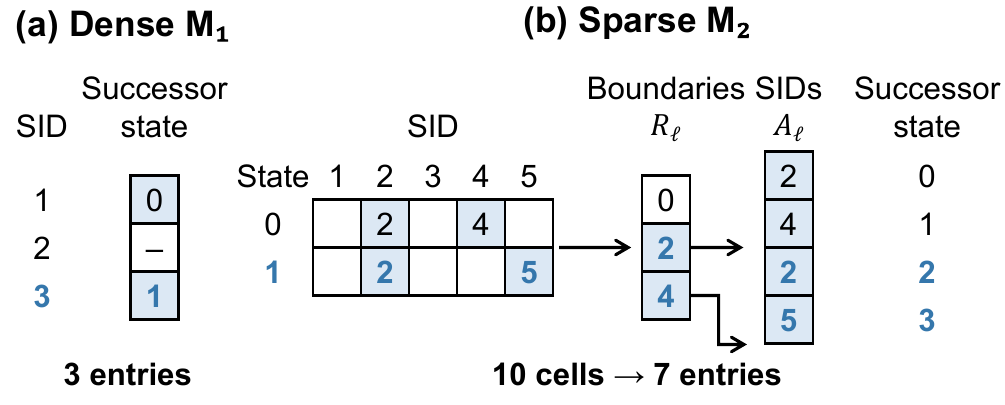}
  \vspace{-4mm}
  \caption{Dense step-$1$ (a) and sparse step-$2$ (b) mappings, where entry positions in (b) serve as successor states.}
  \Description{Panel (a) shows the dense step-one mapping over tokens one to three. Its successor row for state zero is zero,invalid,one. The cell of token three, which holds successor state one, is outlined, and an arrow leads from it to state one in panel (b). Panel (b) shows the sparse step-two mapping with boundaries zero,two,four and tokens two,four,two,five. Entry positions zero through three give the successor states without a stored array; token five at position three and successor three are outlined.}
  \label{fig:constraint-index}
  \vspace{-4mm}
\end{figure}

Figure~\ref{fig:constraint-index} shows the dense and sparse mappings of the first two steps of Figure~\ref{fig:step-sliced-index} and traces the items that begin with $(3,5)$.
The dense step-$1$ mapping has one column for each of the tokens $1$ to $3$; the cell of token~$2$ holds the sentinel, and the cell of token~$3$ holds state~$1$, the step-$2$ state of prefix $(3)$.
In the sparse step-$2$ mapping, boundaries $2$ and $4$ delimit the segment of state~$1$, which lists the valid tokens $2$ and $5$.
Token~$5$ lies at position~$3$, which is also the step-$3$ state of prefix $(3,5)$.
The lookup thus crosses from a dense to a sparse mapping through a single state.

Besides supplying candidates in the form each step's projection needs, the representation is designed to keep the memory footprint of the index small~\citep{zhang2018surf,binna2018hot}, because the index resides in GPU memory alongside the model weights and attention caches.
Its size grows with the number of valid extensions and states it represents, which increase toward the last step: the extensions at step~$\ell$ correspond one-to-one to the distinct prefixes of length~$\ell$, and at step~$L$ these prefixes are the SID sequences of all items.
The step-sliced index reduces this storage in two ways: sparse mappings store no successor states for their extensions, and completed SID sequences need no storage beyond their final SIDs.
First, the position of each entry in a sparse mapping serves as its successor state.
Each prefix at step~$\ell+1$ arises from exactly one extension at step~$\ell$, so numbering these extensions by position gives every such prefix a distinct state, and $\mathcal{M}_{\ell+1}$ numbers its rows the same way.
A sparse mapping thus needs no successor array, which would be as long as its SID array.

Second, every SID sequence has exactly $L$ tokens, so the step number alone marks an SID selected at step~$L$ as the last of its sequence, and the index needs no field that records completion.
$\mathcal{M}_L$ therefore lists the valid final SIDs of each prefix without successor states.
The index also stores no item identifiers, because a completed SID sequence already identifies its item, so the completed sequences, one per item and thus the largest set of prefixes, need no storage beyond their final SIDs.

\subsection{Execution Model}
\label{sec:index-execution}

With the step-sliced index resident in GPU memory, \sys processes online requests by running each decoding step of Figure~\ref{fig:zerogr-overview} on the GPU.
At every step, \sys filters the candidates of each beam through the index before normalizing their logits into scores, and this order also allows it to compute only the logits of the valid tokens.
However, valid tokens differ in number from beam to beam and from step to step, so computing the logits of exactly these tokens is irregular work~\citep{hong2024themis}, which a GPU executes less efficiently than the regular matrix multiplication of full projection.
Filtering, projection, top-$k$ selection, and gather all handle these valid tokens, so the time saved by computing fewer logits also depends on how the stages pass them to one another.
The rest of this subsection therefore describes \emph{shared-slot execution}, which organizes these hand-offs around candidate slots that all stages of a step share, and \emph{hybrid projection}, which chooses full or selective projection for each step.
It describes these slots, how filtering and projection fill them, and how top-$k$ selection and gather read them.

\subsubsection{Shared Candidate Slots}
\label{sec:stage-interface}

To execute the stages of a decoding step efficiently on the GPU, \sys gives every beam of the step the same number of \emph{slots}, which all stages share.
Each candidate, a beam extended by one of its valid tokens, occupies one slot of its beam: filtering and projection write the candidate into this slot, and selection and gather find it by the slot's position alone.

Specifically, at step~$\ell$, the slots of each beam form one row, and all rows of the step have the same length: the length of the step's longest segment for a sparse mapping, or $|\mathcal{V}_\ell|$ for a dense mapping.
Each slot holds the token, successor state, and score of one candidate.
With a sparse mapping, the beam's valid tokens fill the first slots of its row in the order of its segment; with a dense mapping, the slots are aligned with the columns of the successor table.
A slot without a valid token holds the score $-\infty$, which excludes it from normalization and selection.
Since all rows have the same length, the position of a slot depends only on its beam and its offset within the row, not on how many candidates the other beams have.

\begin{figure}[t]
  \centering
  \includegraphics[width=0.8\linewidth]{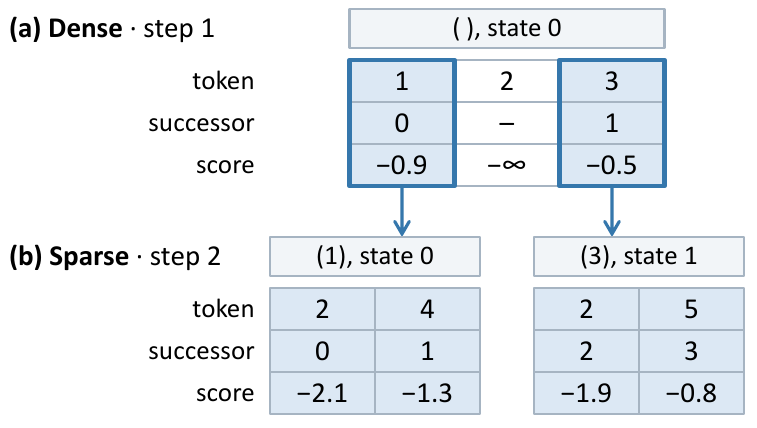}
  \vspace{-4mm}
  \caption{Shared candidate slots in the first two steps of the index, with $B=2$.}
  \Description{Two panels on the index whose SID sequences are (1,2,5), (1,2,7), (1,4,3), (3,2,6), (3,5,1), and (3,5,4). Panel (a), dense mapping at step 1: the beam with the empty prefix and state 0 has three slots for tokens 1, 2, and 3, with successor states 0, none, and 1 and scores -0.9, minus infinity, and -0.5. The slots of tokens 1 and 3 are outlined as selected, and arrows lead to the step-2 beams (1) with state 0 and (3) with state 1. Panel (b), sparse mapping at step 2: these beams have slots for tokens 2 and 4 with successor states 0 and 1, and for tokens 2 and 5 with successor states 2 and 3; their scores are -2.1, -1.3, -1.9, and -0.8.}
  \label{fig:candidate-slots}
  \vspace{-4mm}
\end{figure}

Figure~\ref{fig:candidate-slots} shows an example of the shared candidate slots for the first two steps of the index in Figure~\ref{fig:step-sliced-index}.
Step~$1$ has a dense mapping, so its single row has a slot for each of the tokens $1$ to $3$, and the slot of token~$2$, whose successor-table cell holds the sentinel, has the score $-\infty$.
Step~$2$ has a sparse mapping, so each beam's row holds the tokens of its segment in order: beam $(1)$, in state~$0$, holds tokens~$2$ and~$4$, and beam $(3)$, in state~$1$, holds tokens~$2$ and~$5$.
The successor state of each slot is the position of its entry in the sparse mapping, so the four slots hold states $0$, $1$, $2$, and~$3$, which identify $(1,2)$, $(1,4)$, $(3,2)$, and $(3,5)$ at step~$3$.

\begin{figure}[t]
  \centering
  \includegraphics[width=\linewidth]{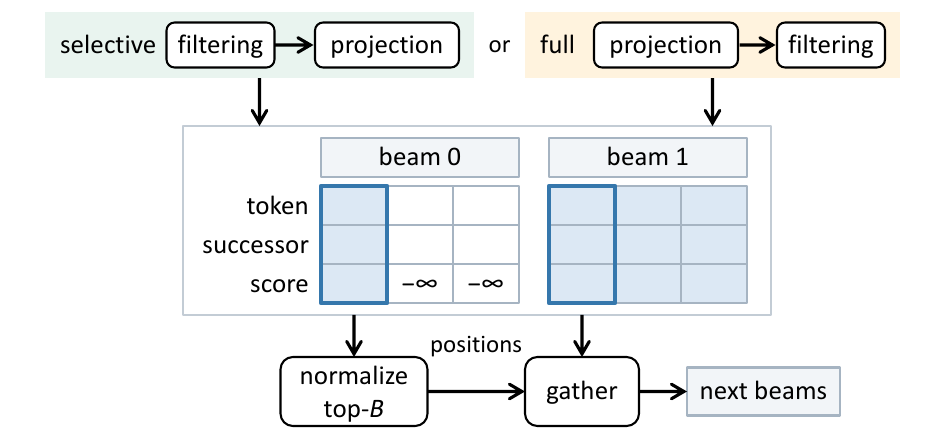}
  \vspace{-4mm}
  \caption{Shared candidate slots, filled by filtering and projection in either order and read by selection and gather.}
  \Description{Two alternatives at the top, selective projection (filtering, then projection) and full projection (projection, then filtering), both lead into one table of candidate slots. The table has a group of three slots for beam 0 and for beam 1, with rows for token, successor, and score. Beam 0 has one valid slot followed by two empty slots whose scores are minus infinity; all three slots of beam 1 are valid. The first slot of each beam is outlined as selected. Below the table, normalize and top-B read the slots and pass positions to gather, which also reads the slots and produces the next beams.}
  \label{fig:shared-slots}
  \vspace{-4mm}
\end{figure}

The shared candidate slots are filled by filtering and projection, which can run in either order (Figure~\ref{fig:shared-slots}).
With a sparse mapping, the order depends on the projection mode.
In selective projection, filtering first expands each beam's segment into its row, writing the tokens and successor states and setting every logit to $-\infty$, and projection then computes only the logits of the valid slots and writes them in place.
In full projection, the step first computes the logits of every token of $\mathcal{V}_\ell$, and filtering then expands the segments in the same way but copies each slot's logit from these token-ordered logits.
With a dense mapping, the token-ordered logits form the slots, and filtering only sets the logits of invalid tokens to $-\infty$.
In every case, filtering completes before selection normalizes the logits into scores, so both projection modes filter first and differ only in whether they compute the logits of invalid tokens.

Once filled, the slots are read in place by selection and gather.
Selection first normalizes each row, which holds the candidates of one beam, by applying a log-softmax over its valid slots and adding the beam's accumulated score.
It then keeps the $B$ highest-scoring candidates among the rows of each request and returns the positions of their slots.
Gather turns each returned position into a beam of the next step.
Since all rows have the same length, dividing the position by the row length gives the parent beam, and the remainder gives the slot within its row.
Because no slot position depends on the candidates of other beams, no stage has to count the candidates of all beams and compute prefix sums over these counts before it places or finds a candidate, and the stages run back to back at every hand-off.

The slot layout strikes a balance between giving every token of $\mathcal{V}_\ell$ a slot and packing only the valid candidates~\citep{billeter2009compaction}.
Sizing each row by the step's longest segment keeps positions fixed with the least padding, and the next two subsections describe how projection and selection further limit the work spent on empty slots.

\subsubsection{Full and Selective Projection}
\label{sec:projection-modes}

Full and selective projection both aim to fill the shared candidate slots as quickly as possible, but they suit steps with different numbers of candidates.

Full projection is the projection of an unconstrained decoder: it computes the logits $z_{\ell,b}=W_\ell h_{\ell,b}$ of all tokens in $\mathcal{V}_\ell$ for all beams in one matrix multiplication~\citep{lhostis2016vocabulary}.
Each row $w_v$ of $W_\ell$ corresponds to a token $v$, and $w_v^{\top} h_{\ell,b}$ is the logit of $v$ for beam~$b$.
This matrix multiplication does not depend on the index, so full projection runs before filtering.
Filtering then builds the slots from the index and fills each valid slot with the logit of its token, while all other slots hold $-\infty$ and are thus excluded when selection normalizes the logits into scores.
Full projection computes many logits that are then discarded, but its matrix multiplication is regular, which is GPU friendly~\citep{grave2017adaptive}.
Its cost depends on $|\mathcal{V}_\ell|$ and the number of beams, but not on how many tokens are valid.

Selective projection computes only the logit $w_v^{\top} h_{\ell,b}$ of each valid slot.
Therefore, it runs after filtering has found the valid tokens.
The arithmetic of selective projection is proportional to the number of candidates, but a direct implementation does not reduce latency in the same proportion.
If it assigns GPU threads to every slot, most of them are idle at later steps, where most slots are empty, and every candidate reloads its beam's hidden vector.

\begin{figure}[t]
  \centering
  \includegraphics[width=\linewidth]{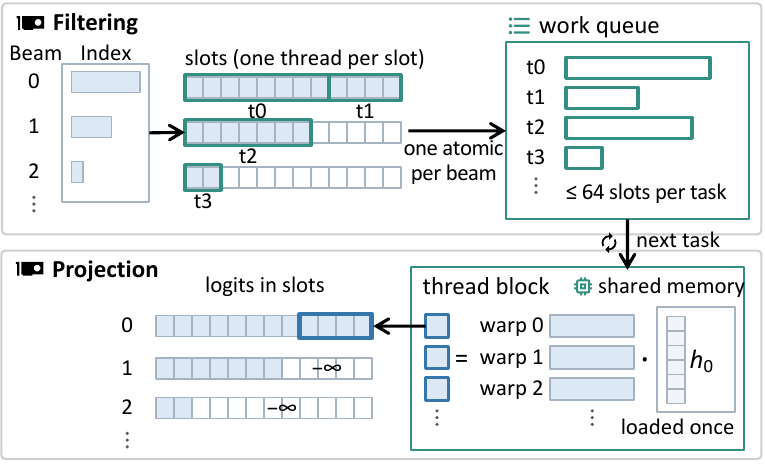}
  \vspace{-6mm}
  \caption{GPU execution of selective projection.}
  \vspace{-4mm}
  \Description{Two panels. In the filtering kernel panel, an index holds one segment for each of beams 0, 1, 2, and more, and an arrow leads to the slots, one row per beam, labeled one thread per slot; each row holds valid slots up to the length of its segment and empty slots after them. Green outlines group the valid slots into tasks t0 to t3: beam 0 splits into t0 and t1, beam 1 forms t2, and beam 2 forms t3. An arrow labeled one atomic per beam leads to a work queue that lists tasks t0 to t3 and more as empty green frames of the same lengths, noting at most 64 slots per task. In the projection kernel panel, an arrow labeled next task leads from the queue to a thread block whose shared memory holds the hidden vector h0, loaded once, and warps 0, 1, 2, and more multiply weight rows with h0 to produce a column of logits. An arrow writes these logits into the slots of task t1 in the logits in slots, where the empty slots hold minus infinity.}
  \label{fig:selective-projection}
  \vspace{-4mm}
\end{figure}

To leverage the benefits of shared candidate slots, \sys designs a dedicated filtering kernel and a selective projection kernel, as shown in Figure~\ref{fig:selective-projection}.
The filtering kernel builds the slots from the index together with a work queue that lists only the valid slots, and the projection kernel then computes the logits of these slots from the queue.
Specifically, the filtering kernel assigns one thread to each slot.
The thread for the $k$-th slot of a beam copies the $k$-th entry of the corresponding segment, a valid token together with its successor state, into the slot, and leaves the slot empty if the segment has fewer than $k$ entries.
Every thread also sets the logit of its slot to $-\infty$, which the projection kernel later overwrites for the valid slots.
Alongside the slots, the filtering kernel builds a work queue whose tasks each cover a run of consecutive valid slots of one beam, up to a fixed length.
The queue lets the projection kernel skip empty slots without moving any candidate out of its slot.

The projection kernel consumes the queue with a fixed number of thread blocks, which loop over it until every task is processed.
For each task, a block first loads the hidden vector of the corresponding beam into on-chip shared memory, and its warps (groups of 32 threads) then process the candidates of the task in parallel, one candidate per warp at a time.
A warp computes the logit of its candidate token as the dot product of the shared hidden vector with the weight row of that token, splitting the hidden dimension across its threads and summing their partial products with a warp-level reduction.
The warp then writes the logit directly into the slot that holds the candidate, where selection later reads it.
Empty slots thus receive no work and keep the $-\infty$ that the filtering kernel wrote, each hidden vector is read once per task rather than once per candidate, and a beam with many valid tokens yields several tasks that can run on different blocks.

Since neither projection mode is faster at every step, \sys realizes hybrid projection by selecting one of three combinations of mapping and projection mode for each step: a dense mapping with full projection, or a sparse mapping with either full or selective projection.
At each model update, before the new version accepts traffic, \sys runs warm-up requests under every combination available to a step and keeps the one with the lowest measured latency.
The selected combinations then stay fixed until the next update, so the decoding loop makes no per-request decision.

\subsubsection{Top-\texorpdfstring{$k$}{k} Selection and Gather}
\label{sec:selection-gather}

Selection turns each beam's logits into scores with a log-softmax over its valid slots, adds the beam's accumulated score, and keeps the $B$ highest-scoring candidates of each request.
A request has $\rho_\ell B|\mathcal{V}_\ell|$ candidates, a number that varies widely across steps: it is a sizable fraction of $B|\mathcal{V}_\ell|$ at early steps but barely exceeds $B$ at later steps, where most prefixes admit only one or two tokens.
\sys therefore performs this selection with two top-$k$ kernels.
For many candidates, a radix-selection kernel~\citep{shanbhag2018topk} splits the slots of a request across many thread blocks and finds the $B$-th highest score in several passes over them, and a separate kernel then performs gather.
When the candidates of a request fit in one thread block, a small top-$k$ kernel instead runs the whole selection in that block: the block locates the valid slots from the segment length of each beam and sorts only their scores in shared memory.
Because the position of a selected slot determines the next beam, the same block also performs gather, so selection and gather take a single kernel.
Both kernels select the same candidates, so the choice between them affects only latency.

Gather forms the beams of the next step from the selected slots.
Each new beam extends the prefix of its parent beam by the token of its slot.
As with the mapping and projection mode, \sys fixes the top-$k$ kernel of each step, and thus whether gather is fused with selection, during warm-up.

\subsection{Index Construction}
\label{sec:cpu-staged-construction}

With the representation of each mapping fixed, \sys must rebuild the index within the update window each time the item catalog is refreshed.
The rebuild runs while the instance continues to serve online requests, so it must not compete with inference for GPU compute or host-to-GPU (H2D) bandwidth.
To meet these requirements, \sys introduces \emph{layout-guided streaming construction}.

In order to meet the online serving requirements, the index is constructed in two CPU passes: the first determines the output layout, and the second generates the mappings in parallel while streaming them to the GPU.
The input is a host-resident \emph{item SID table} $\mathcal{T}$ that lists every item with its length-$L$ SID sequence, sorted lexicographically by sequence.
We number the items in this order, so the $i$-th row holds $\mathbf{s}(i)$.
The output is the GPU-resident mappings $\mathcal{M}_1,\ldots,\mathcal{M}_L$, each in its chosen representation.

\begin{algorithm}[t]
\caption{Two-pass construction of sparse mappings}
\label{alg:index-construction}
\begin{algorithmic}[1]
\Require Sorted item SID table $\mathcal{T}$
\Ensure Sparse mappings $\mathcal{M}_\ell=(A_\ell,R_\ell)$ for $\ell=1,\ldots,L$
\State $n_\ell\gets0$ for $\ell=1,\ldots,L$
    \label{alg:construction-count-start}
\For{$i=1,\ldots,|\mathcal{T}|$}
    \State $h\gets\operatorname{LCP}(\mathbf{s}(i-1),\mathbf{s}(i))$
    \State $n_\ell\gets n_\ell+1$ for $\ell=h+1,\ldots,L$
\EndFor
    \label{alg:construction-count-end}
\State Allocate $A_\ell$ and $R_\ell$ for every $\ell$ from the counts
    \label{alg:construction-layout}
\For{$i=1,\ldots,|\mathcal{T}|$}
    \label{alg:construction-emit-start}
    \State $h\gets\operatorname{LCP}(\mathbf{s}(i-1),\mathbf{s}(i))$
    \For{$\ell=h+1,\ldots,L$}
        \State Append $s^{(i)}_\ell$ to $A_\ell$
            \label{alg:construction-append}
        \State \textbf{if} $\ell<L$ \textbf{then} append $|A_{\ell+1}|$ to $R_{\ell+1}$
            \label{alg:construction-boundary}
    \EndFor
\EndFor
    \label{alg:construction-emit-end}
\State \Return $\mathcal{M}_1,\ldots,\mathcal{M}_L$
\end{algorithmic}
\end{algorithm}

Algorithm~\ref{alg:index-construction} shows how layout-guided streaming construction builds sparse mappings.
Lines~\ref{alg:construction-count-start}--\ref{alg:construction-count-end} form the first pass, which determines how much space each mapping needs before any SID is written.
Its counter $n_\ell$ records the number of valid continuations at step~$\ell$, which is the number of SIDs that $\mathcal{M}_\ell$ stores.
Because $\mathcal{T}$ is sorted, sequences that share a prefix are adjacent.
A sequence therefore repeats the valid continuations of its predecessor up to their longest common prefix (LCP) and introduces new ones only at the steps beyond it~\citep{daciuk2000automata}.
For an LCP of length $h$, the algorithm increments $n_{h+1},\ldots,n_L$.
The first sequence takes the empty sequence $\mathbf{s}(0)=\epsilon$ as its predecessor, so it introduces a valid continuation at every step.
After the first pass finishes, Line~\ref{alg:construction-layout} allocates the mappings from these counts.
Each $n_\ell$ sizes the SID array $A_\ell$ of $\mathcal{M}_\ell$ and, for $\ell<L$, also the boundary array $R_{\ell+1}$ of $\mathcal{M}_{\ell+1}$, because each valid continuation at step~$\ell$ yields one state at step~$\ell+1$.

Once the arrays are allocated, the second pass in Lines~\ref{alg:construction-emit-start}--\ref{alg:construction-emit-end} scans $\mathcal{T}$ again in the same order and fills them.
For each sequence, Line~\ref{alg:construction-append} appends the SIDs at steps $h+1$ to $L$, which are its new valid continuations, to the corresponding SID arrays.
Line~\ref{alg:construction-boundary} then opens, for every appended SID, the segment that will hold its continuations at the next step, by recording the current end of $A_{\ell+1}$ in $R_{\ell+1}$.
The segments of $\mathcal{M}_{\ell+1}$ are thus created in the same order as the SIDs in $A_\ell$, so the position of each SID directly serves as its successor state.

\begin{figure}[t]
  \centering
  \includegraphics[width=\linewidth]{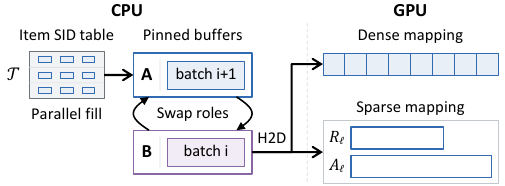}
  \vspace{-6mm}
  \caption{Double-buffered index construction.}
  \Description{The CPU fills one pinned buffer while the other transfers a previous batch to the GPU. The GPU stores a dense array or sparse token and boundary arrays.}
  \vspace{-4mm}
  \label{fig:cpu-staged-construction}
\end{figure}

Both passes are easy to parallelize, because each sequence is compared only with its predecessor.
\sys partitions $\mathcal{T}$ into $C$ contiguous chunks and runs the two passes over the chunks on separate CPU threads.
In the first pass, each chunk counts its own valid continuations at every step, and exclusive prefix sums~\citep{blelloch1990prefix} over these per-chunk counts assign each chunk a disjoint output range at every step.
In the second pass, each chunk then writes its fragment directly into its assigned ranges, independently of the other chunks.
Dense mappings fit into the same passes even easily: instead of appending the SID, the second pass writes the successor state into the cell of that SID.

The two passes establish the placement and contents of the index fragments.
Figure~\ref{fig:cpu-staged-construction} shows their delivery to the GPU through two reusable pinned buffers.
Adjacent chunks form batches that fit a buffer, allowing CPU generation of one batch to overlap asynchronous H2D transfer of another into its final array locations~\citep{nvidiaCUDAguide}.
Double buffering helps the rebuild meet both of its online requirements.
First, it shortens construction: while the CPU fills one buffer, the other is being copied to the GPU, so transfer effectively overlaps with generation.
Second, it avoids a burst of H2D traffic.
Transfer starts as soon as the first batch is ready, so the index reaches the GPU in small batches spread across the rebuild rather than in one large copy at its end.
Inference requests served at the same time therefore do not wait behind a long index transfer.

\section{Evaluation}
\label{sec:evaluation}

This section answers three questions about \sys:
\begin{itemize}
  \item How much does \sys reduce latency relative to the baselines on real-world and synthetic datasets, and which steps and stages account for it (\S~\ref{sec:eval-end-to-end} and \S~\ref{sec:eval-latency})?
  \item How much does each design choice contribute (\S~\ref{sec:eval-ablation})?
  \item How much memory does the \sys index save, and does its construction complete within the update window without competing with online serving for GPU compute or host-to-GPU bandwidth (\S~\ref{sec:eval-memory} and \S~\ref{sec:eval-construction})?
\end{itemize}

\subsection{Experimental Setup}
\label{sec:eval-setup}

We run experiments on a server with one NVIDIA GeForce RTX 4090 D GPU and two Intel Xeon Gold 5520+ processors.
Some experiments are run on a server with a data center GPU.
We compare \sys with four baselines.
\begin{itemize}
  \item \emph{Unconstrained} performs full computation without any constraints.
  It serves as a performance reference; its outputs need not correspond to items in the catalog.
  \item \emph{CPU trie} is the constrained decoder of our previously deployed serving system, which keeps the prefix tree in host memory and retrieves valid continuations on the CPU.
  \item \emph{FlashTrie}~\citep{anandakumar2026flashtrie} encodes the prefix tree as a succinct bit-vector structure in GPU memory and searches it on the GPU.
  \item \emph{STATIC}~\citep{su2026static} flattens the prefix tree into a sparse index and performs constraint filtering in GPU memory.
\end{itemize}
The CPU trie, FlashTrie, and STATIC baselines score all continuations before filtering, whereas \sys filters first and scores only the valid continuations. 
\sys and the CPU trie are implemented in C++ and CUDA, and FlashTrie and STATIC run on PyTorch 2.11.

We evaluate on two real-world datasets and one synthetic dataset.
The \emph{Search} dataset is collected from an industrial generative search service and contains nearly one billion distinct five-token SID sequences.
Its model projects onto a single 18,432-token output head at every step, although the tokens of each step occupy a disjoint band of 8,192, 4,096, 2,048, 2,048, or 2,048 entries.
The \emph{Watch} dataset contains approximately 4.17 billion distinct three-token SID sequences and uses an 8,192-token output head at each step.
The synthetic dataset, \emph{Synthetic-7}, has seven-token SID sequences, and its variants differ in vocabulary size and branching factors.

\subsection{End-to-End Performance}
\label{sec:eval-end-to-end}
In this section, we measure the total latency of all $L$ decoding steps on both real-world and synthetic datasets, excluding model inference time.
We use beam width 128 and request batch sizes 1 to 32, with hidden size 4,096 for FP32 and 1,024 for FP16.

\subsubsection{Real-World Datasets}
\label{sec:eval-end-to-end-real}

\begin{figure}[t]
  \centering
  \includegraphics[width=\linewidth]{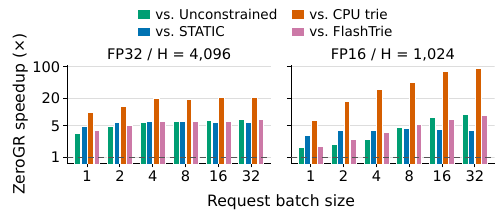}
  \vspace{-8mm}
  \caption{Speedup of \sys over the baselines on the Search dataset, measured on 4090 D.}
  \Description{The left panel uses FP32 with hidden size 4,096, and the right panel uses FP16 with hidden size 1,024. Both panels use a logarithmic speedup axis with a dashed reference line at one.}
  \label{fig:search-end-to-end}
  \vspace{-4mm}
\end{figure}

For the real-world datasets, we evaluate \sys on the Search and Watch datasets with two environments: 4090 D and a data center GPU.
To evaluate \sys performance on a real-world workload across batch sizes and precisions, Figure~\ref{fig:search-end-to-end} shows its speedup over the baselines on the Search dataset, measured on 4090 D.
\sys outperforms all baselines at every batch size and precision, including Unconstrained, which enforces no constraints.
At batch size 32, \sys completes all five decoding steps in 9.68\,ms with FP32 and 1.88\,ms with FP16, the latter $9.1\times$ faster than Unconstrained.
With FP16, the speedup over Unconstrained is smaller at small batch sizes, where the hidden size of 1,024 already makes full projection inexpensive.
The gap to the CPU trie widens with the batch size and approaches two orders of magnitude with FP16, because its host-side lookup grows with the number of active beams.

At large batch sizes, STATIC is even faster than Unconstrained.
This is because we adapt STATIC to apply the softmax only over the valid tokens, as \sys does, whereas Unconstrained normalizes over all 18,432 tokens, and the smaller softmax can outweigh the lookup cost of STATIC as the batch grows.
These results show that applying the constraints before projection reduces decoding work instead of adding to it.
On the Search dataset, every beam is projected onto a single 18,432-token output head at each step, although only a small fraction of these tokens can extend the prefix of the beam.
Unconstrained and STATIC compute the logits of all these tokens, and STATIC uses the constraints only to mask the logits afterward.
\sys instead filters the candidates of each beam first and, at the sparse later steps, computes only the logits of the valid tokens, which removes most of the projection work.

\begin{figure}[t]
  \centering
  \includegraphics[width=\linewidth]{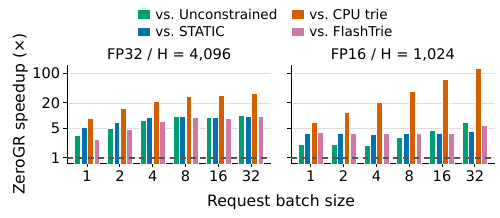}
  \vspace{-8mm}
  \caption{Speedup of \sys over the baselines on the Search dataset, measured on the data center GPU.}
  \Description{The left panel uses FP32 with hidden size 4,096, and the right panel uses FP16 with hidden size 1,024. Both panels share a logarithmic speedup axis with a dashed reference line at one. Each speedup divides the baseline's uninstrumented whole-batch primary CUDA-event median by ZeroGR's median.}
  \label{fig:search-end-to-end-h200}
  \vspace{-4mm}
\end{figure}

Figure~\ref{fig:search-end-to-end-h200} further shows that \sys outperforms the baselines on the Search dataset when measured on the data center GPU.
With FP16, the latency of \sys stays between 0.93 and 1.07\,ms from batch size 1 to 32, whereas Unconstrained and the CPU trie slow down as the batch size grows.
Computing only valid logits limits the latency increase for larger batches, giving \sys its greatest advantage at the large batch sizes used in production.

\begin{figure}[t]
  \centering
  \includegraphics[width=\linewidth]{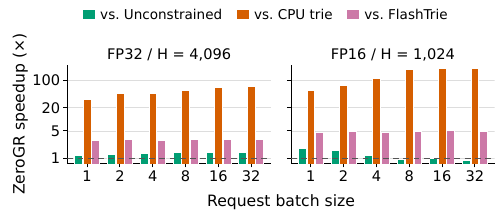}
  \vspace{-8mm}
  \caption{Speedup of \sys over the baselines on the Watch dataset, measured on 4090 D.}
  \vspace{-4mm}
  \Description{The left panel uses FP32 with hidden size 4,096 and shows speedups over the CPU trie from 32.4 to 68.7 times and over Unconstrained from 1.21 to 1.43 times. The right panel uses FP16 with hidden size 1,024 and shows speedups over the CPU trie from 54.0 to 204.9 times and over Unconstrained from 0.91 to 1.82 times. Both panels share a logarithmic speedup axis with a dashed reference line at one; values below one are retained.}
  \label{fig:watch-end-to-end}
\end{figure}

To evaluate \sys on a catalog in which most tokens remain valid, Figure~\ref{fig:watch-end-to-end} shows its speedup over the baselines on the Watch dataset, measured on 4090 D.
STATIC runs out of GPU memory on this GPU and is therefore omitted.
\sys outperforms the CPU trie by up to $205\times$, because the CPU trie returns thousands of valid tokens per beam from the host at the dense steps.
Relative to Unconstrained, \sys is faster at every batch size with FP32 and at small batch sizes with FP16.
These results show that \sys guarantees valid outputs at almost the speed of unconstrained decoding, and often faster, even when the constraints remove little computation.
On the Watch dataset, every token is valid at the first step and about 90\% of the tokens remain valid at the second step, so only the last of the three steps is sparse.
\sys uses full projection for most of the dense steps and selective projection at the sparse last step, so it adds no projection work where the constraints remove little and saves work where they remove much.
Its small slowdown with FP16 at large batch sizes is expected, since Unconstrained skips the constraint lookup entirely and does not guarantee that its outputs correspond to items in the catalog.

\begin{figure}[t]
  \centering
  \includegraphics[width=\linewidth]{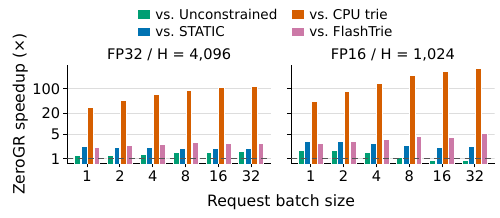}
  \vspace{-8mm}
  \caption{Speedup of \sys over the baselines on the Watch dataset, measured on the data center GPU.}
  \vspace{-4mm}
  \Description{The left panel uses FP32 with hidden size 4,096, and the right panel uses FP16 with hidden size 1,024. Both panels share a logarithmic speedup axis with a dashed reference line at one; values below one are retained. Each speedup uses uninstrumented whole-batch primary CUDA-event medians.}
  \label{fig:watch-end-to-end-h200}
\end{figure}

Figure~\ref{fig:watch-end-to-end-h200} further shows the results on the Watch dataset when measured on the data center GPU, whose memory is large enough for STATIC.
Although the Watch dataset is less favorable to \sys than the Search dataset, because its constraints remove little projection work, \sys remains $2.0$--$3.2\times$ faster than STATIC.
Its speedups over Unconstrained and the CPU trie follow the same trends as on 4090 D.
The per-step choice of projection mode keeps these trends stable across GPUs: \sys uses full projection at the dense steps and selective projection where it is faster, and its choices adapt to the GPU, precision, and batch size.
Since STATIC is slower than Unconstrained at every batch size, applying the constraints after full projection adds substantial overhead when most tokens are valid, whereas \sys stays close to Unconstrained.

\subsubsection{Synthetic Dataset}
\label{sec:eval-end-to-end-synthetic}

Real-world datasets fix the branching factor at each step and the width of the output heads.
We vary these factors through the variants of Synthetic-7.

We evaluate all variants on 4090 D.
Synthetic-7 contains $2^{30}$ distinct seven-token SID sequences; all seven steps share a $V$-token output head, and its variants differ in $V$.

\begin{figure}[t]
  \centering
  \includegraphics[width=\linewidth]{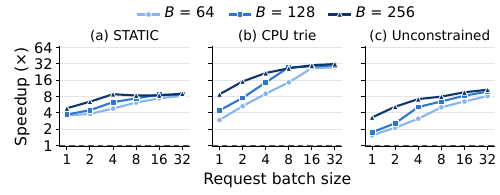}
  \vspace{-8mm}
  \caption{Speedup of \sys over the baselines on \mbox{Synthetic-7} for beam widths $B$ from 64 to 256.}
  \vspace{-2mm}
  \Description{Three side-by-side line charts plot speedup against request batch sizes 1, 2, 4, 8, 16, and 32 on a shared logarithmic axis from 1 to 64, with a dashed reference line at one. Panel a shows ZeroGR speedup over STATIC, from 3.6 to 9.0 times; panel b shows speedup over the CPU trie, from 3.0 to 31.7 times; panel c shows speedup over Unconstrained, from 1.54 to 10.67 times. Each panel has one line per beam width of 64, 128, and 256, drawn in progressively darker blue with distinct markers. Speedup is baseline median synchronized wall-clock latency divided by fixed-v10 ZeroGR median latency. All seven SID positions use an 8,192-token vocabulary.}
  \label{fig:synthetic7-beam}
\end{figure}

To evaluate how \sys scales with the number of beams, Figure~\ref{fig:synthetic7-beam} shows its speedup on Synthetic-7 with FP32 for beam widths from 64 to 256 and $V=8{,}192$.
The speedup over every baseline grows with the batch size and is generally higher for wider beams, reaching $10.7\times$ over Unconstrained at beam width 256 and batch size 32, where \sys takes 8.28\,ms.
These results show that \sys scales better than the baselines as a batch contains more beams.
Full projection computes $V$ logits for every beam, so the projection cost of Unconstrained and STATIC grows with the total number of beams.
\sys instead chooses the projection mode for each step, improving scaling as the number of beams grows.

\begin{figure}[t]
  \centering
  \includegraphics[width=\linewidth]{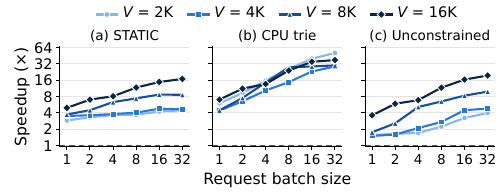}
  \vspace{-6mm}
  \caption{Speedup of \sys over the baselines on \mbox{Synthetic-7} for vocabulary sizes $V$ from 2,048 to 16,384.}
  \vspace{-2mm}
  \Description{Three side-by-side line charts plot speedup against request batch sizes 1, 2, 4, 8, 16, and 32 on a shared logarithmic axis from 1 to 64, with a dashed reference line at one. Panel a shows ZeroGR speedup over STATIC, from 3.3 to 16.9 times; panel b shows speedup over the CPU trie, from 4.4 to 50.9 times; panel c shows speedup over Unconstrained, from 1.50 to 19.53 times. Each panel has one line per vocabulary size of 2,048, 4,096, 8,192, and 16,384, drawn in progressively darker blue with distinct markers. Speedup is baseline median synchronized wall-clock latency divided by fixed-v10 ZeroGR median latency. Catalog cardinality remains 2 to the power of 30, while the first- and final-step branching factors vary with vocabulary size.}
  \label{fig:synthetic7-vocab}
  \vspace{-2mm}
\end{figure}

To evaluate how the width of the output head affects \sys, Figure~\ref{fig:synthetic7-vocab} varies $V$ from 2,048 to 16,384 with beam width 128.
The speedup over Unconstrained and STATIC generally grows with $V$, reaching $19.5\times$ over Unconstrained at $V=16{,}384$ and batch size 32, while the latency of \sys at batch size 32 stays between 3.1 and 4.5\,ms across all four vocabulary sizes.
These results show that the cost of \sys follows the number of valid tokens rather than the width of the output head.
As $V$ grows, full projection computes more logits for every beam, whereas the number of valid next tokens stays at four for steps~2 to~6 and even decreases at step~7.
This property allows production deployments to adopt larger codebooks, for example to represent a growing item catalog, without a corresponding increase in decoding latency.

\subsection{Latency Breakdown}
\label{sec:eval-latency}

\begin{figure}[t]
  \centering
  \includegraphics[width=\linewidth]{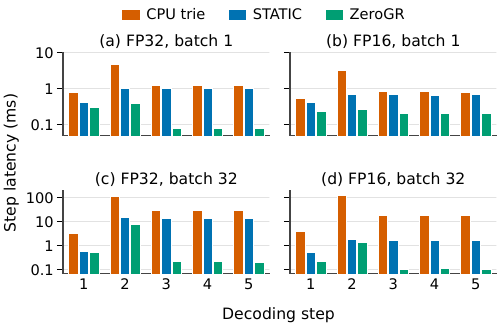}
  \vspace{-6mm}
  \caption{Per-step latency of \sys, the CPU trie, and STATIC on the Search dataset.}
  \Description{Four panels in a two-by-two grid show median latency for decoding steps one through five. The top row shows FP32 batch 1 and FP16 batch 1; the bottom row shows FP32 batch 32 and FP16 batch 32. Every panel has grouped bars for the CPU trie, STATIC, and ZeroGR on a logarithmic latency axis, with a shared legend and matching latency ranges within each row.}
  \vspace{-4mm}
  \label{fig:search-step-latency}
\end{figure}

To understand the sources of \sys's performance improvements, Figure~\ref{fig:search-step-latency} breaks the latency on the Search dataset down by decoding step.
Step~2 is the most expensive step of \sys in every configuration, and the latency of \sys drops sharply after it: with FP32 at batch size 1, each of steps~3 to~5 takes about 0.08\,ms, compared with about 1.04\,ms for STATIC.
This pattern follows from the design of \sys, which computes only the logits of valid tokens wherever this is cheaper than full projection, so the cost of each step follows how many tokens are valid at that step.
Step~2 has the most valid tokens of all steps, whereas few tokens are valid at the later steps, which therefore cost almost nothing; the baselines, in contrast, pay for the full output head at every step.

\begin{figure}[t]
  \centering
  \includegraphics[width=\linewidth]{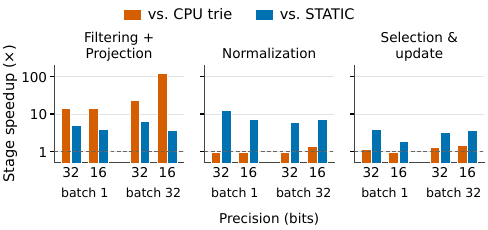}
  \vspace{-6mm}
  \caption{Per-stage speedup of \sys over the CPU trie and STATIC on the Search dataset.}
  \Description{Three panels in one row show speedup for combined filtering and projection, normalization, and selection and update. Filtering includes gather. The combined-stage speedup divides the sum of a baseline's filtering and projection times by the corresponding sum for ZeroGR; all stage times are summed over the five decoding steps. Each panel groups precision values of 32 and 16 bits by request batch sizes 1 and 32, with orange bars for speedup over the CPU trie and blue bars for speedup over STATIC. All panels share a logarithmic speedup axis with a dashed reference line at one.}
  \vspace{-2mm}
  \label{fig:search-stage-breakdown}
\end{figure}

To see where the time goes within each step, Figure~\ref{fig:search-stage-breakdown} reports the speedup of \sys for three stages: filtering and projection, normalization, and selection and update.
Filtering and projection form the most expensive stage: at batch size 32 with FP32, they account for 97\% of the latency of STATIC and nearly all of the latency of the CPU trie.
\sys gains the most in this stage, up to $120\times$ over the CPU trie, because filtering reads the valid tokens of each beam directly from the per-step mapping on the GPU and selective projection then computes only the logits of these tokens.
Since this stage dominates the decoding latency, computing only valid logits accounts for most of the overall speedup of \sys.
\sys also remains faster than STATIC in the two later stages in every configuration; in \sys, both stages read the shared candidate slots in place, and selection uses the top-$k$ kernel chosen for each step.
Against the CPU trie, these two stages perform about the same, since the CPU trie also runs them on the GPU over the valid candidates.
Thus, the most expensive stage provides most of the speedup of \sys, which the later stages preserve.

\subsection{Ablation Study}
\label{sec:eval-ablation}

Most of the speedup of \sys comes from computing only the logits of valid tokens, but computing fewer logits does not by itself reduce latency.
Table~\ref{tab:ablation-projection} examines this effect at decoding step~2 of the Search dataset, the step with the most valid tokens, with FP32 and batch size 1.
It compares the selective projection of \sys with full projection and with a naive selective projection that computes the same logits without shared candidate slots.
The naive filtering packs the candidates of all beams into a compact list, which requires counting them and computing prefix sums, and the naive projection creates a temporary matrix for valid projection lines before computing the logits in a separate kernel.
Although naive selective projection computes only a small fraction of the logits of full projection, it is $5.1\times$ slower, because it lacks the support of efficient memory access patterns.
\sys avoids these overheads: its filtering kernel writes each candidate into a fixed slot without prefix sums, and its projection kernel reads the weight rows in place and loads each hidden vector once per task.
It thereby computes the same logits $8.5\times$ faster than naive selective projection and also outperforms full projection.
The execution model of \sys is thus what turns the reduction in logits into a reduction in latency.

\begin{table}[t]
  \centering
  \caption{Filtering and projection latency at decoding step~2 on the Search dataset with FP32 and batch size 1.}
  \vspace{-3mm}
  \label{tab:ablation-projection}
  \small
  \begin{tabular*}{\linewidth}{@{\extracolsep{\fill}}lrr@{}}
    \toprule
    Method & Logits & Latency ($\mu$s) \\
    \midrule
    Full projection & 2,359,296 & 507.8 \\
    Naive selective & 71,189 & 2,594.9 \\
    \sys selective & 71,189 & 305.1 \\
    \bottomrule
  \end{tabular*}
\end{table}

\begin{figure}[t]
  \centering
  \captionsetup[subfigure]{font=footnotesize,skip=1pt}
  \begin{subfigure}[t]{0.49\linewidth}
    \centering
    \includegraphics[width=\linewidth]{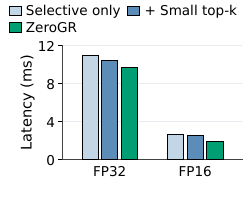}
    \vspace{-6mm}
    \caption{Decoding latency.}
  \end{subfigure}\hfill
  \begin{subfigure}[t]{0.49\linewidth}
    \centering
    \includegraphics[width=\linewidth]{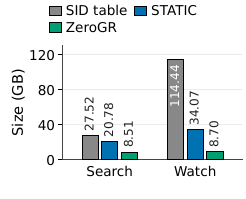}
    \vspace{-6mm}
    \caption{Index memory footprint.}
  \end{subfigure}
  \vspace{-4mm}
  \caption{Decoding latency on the Search dataset at batch size 32 and index memory footprint.}
  \Description{Two subfigures. Panel a groups FP32 and FP16 decoding latency at batch size 32 for selective only, plus small top-k, and ZeroGR. The FP32 latencies are 10.99, 10.47, and 9.68 ms; the FP16 latencies are 2.61, 2.50, and 1.88 ms. Panel b compares SID table, STATIC, and ZeroGR sizes on Search and Watch. On Search the sizes are 27.52, 20.78, and 8.51 GB; on Watch they are 114.44, 34.07, and 8.70 GB.}
  \vspace{-4mm}
  \label{fig:ablation-memory}
\end{figure}

With selective projection in place, \sys further adds a small top-$k$ kernel and selects the projection mode of each step.
Figure~\ref{fig:ablation-memory}(a) adds these two design choices one at a time to a variant that uses selective projection at every step, on the Search dataset at batch size 32.
The small top-$k$ kernel accelerates the later steps, where few candidates remain, by running the selection and gather of each request in one thread block, and reduces the latency of steps~3 to~5 by up to 21\%.
Selecting the projection mode then picks the faster mode for every step and precision: \sys uses full projection at step~1, where almost every token is valid, and at step~2 uses selective projection with FP32 and full projection with FP16.
With FP16, the two design choices together reduce the decoding latency of \sys by 27\% relative to selective projection alone.

\subsection{Memory Footprint}
\label{sec:eval-memory}

To evaluate the memory usage of \sys, Figure~\ref{fig:ablation-memory}(b) compares the GPU-resident indices of STATIC and \sys on the Search and Watch datasets, together with the SID table, stored as a TSV file, from which both indices are built.
The index of \sys is 59.1\% and 74.5\% smaller than that of STATIC on the two datasets, respectively, and it is smaller than the SID table on both.

This reduction comes from how each index stores a valid extension.
STATIC stores a successor state alongside the SID of every extension, whereas \sys stores only the SID, because the position of an entry serves as its successor state and a final SID needs none.
The reduction is larger on the Watch dataset, where final SIDs make up almost the entire index.
On 4090 D, which has 24\,GB of memory, the index of STATIC occupies most of the memory on the Search dataset and does not fit on the Watch dataset, whereas the index of \sys fits on both and leaves most of the memory to the model weights and attention caches.

\subsection{Index Construction Performance}
\label{sec:eval-construction}

\begin{figure}[t]
  \centering
  \includegraphics[width=\linewidth]{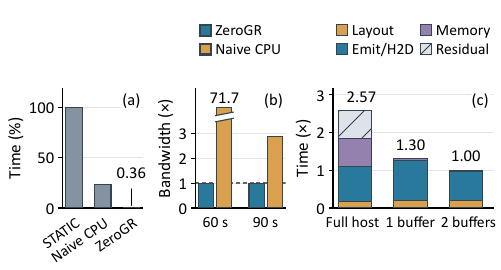}
  \vspace{-6mm}
  \caption{Index construction on the Search dataset. (a) Time relative to STATIC. (b) H2D bandwidth relative to \sys in each window. (c) Staging time relative to \sys.}
  \vspace{-6mm}
  \Description{Panel a compares total construction time relative to STATIC. Panel b groups ZeroGR and Naive CPU by update window, with bandwidth normalized to ZeroGR separately within each window. The broken bar represents 71.7 times ZeroGR's bandwidth under the 60-second window. Panel c shows stage times normalized to ZeroGR's total.}
  \label{fig:construction-performance}
\end{figure}

We evaluate how efficiently \sys rebuilds its filtering index by comparing construction time and H2D requirements on the Search dataset.
Figure~\ref{fig:construction-performance} reports the results, with construction measured from the sorted item SID table in host memory to the GPU-resident mappings.
Alongside STATIC, we include \emph{Naive CPU}, a serial builder of the same step-sliced filtering index as \sys.
It builds the mappings one step at a time without streaming.

Figure~\ref{fig:construction-performance}(a) gives the total construction time of each method relative to STATIC.
\sys completes the rebuild in 0.914\,s, which is 0.36\% of STATIC's time, whereas STATIC exceeds the online service update windows.
Naive CPU takes less time than STATIC, but it remains far slower than \sys even though both build the same index.

The colored bars in Figure~\ref{fig:construction-performance}(b) estimate the H2D bandwidth needed within each update window after CPU construction.
For each window, the bars are normalized to \sys's bandwidth, with the dashed line marking the unit baseline.
With its compact payload and short CPU phase, \sys requires 0.144\,GB/s for a 60\,s window.
STATIC's CPU phase alone exceeds both windows.
Naive CPU uses almost the entire 60\,s window for construction, so it needs very high bandwidth to transfer the index in the remaining time, as the broken bar shows.
Even with the 90\,s window, Naive CPU still requires $2.9\times$ the bandwidth of \sys.

Figure~\ref{fig:construction-performance}(c) compares three variants of the same parallel two-pass builder, with all stage times normalized to \sys's total.
\emph{Full host} materializes and registers the complete host output before transfer.
\emph{Single buffer} generates and transfers batches through one pinned buffer; \sys uses two such buffers to overlap these operations.
\sys is $2.57\times$ faster than Full host and $1.30\times$ faster than Single buffer.
Full host spends most of its extra time registering the complete output as pinned memory and releasing it after the transfer.
Single buffer removes this cost, but its CPU must wait for each transfer to finish before generating the next batch.
Staging through bounded buffers and overlapping generation with transfer lets \sys avoid both overheads.

\section{Related Work}
\label{sec:related-work}

\paragraph{Traditional filtering methods.}
Many GR systems decode without constraints and handle invalid outputs afterward.
Rewarding valid outputs during reinforcement learning makes invalid SID sequences rarer but cannot rule them out~\citep{zhou2025onerectech}.
Post-generation filtering discards candidates that map to no item or violate business rules once decoding completes~\citep{rajput2023tiger,yang2026adabool,wang2026nezha,pang2025higr}.
Neither approach enforces validity during decoding, so invalid prefixes still occupy beam slots and consume decoding work before they are removed.

\paragraph{Constrained decoding for generative recommendation.}
Constrained decoding enforces validity at every step by looking up the valid tokens of each prefix in an index and masking the scores of invalid tokens~\citep{decao2021genre,bevilacqua2022seal,willard2023outlines,zheng2024lcrec,li2025leadre,sun2025xgr}.
To speed up this per-step lookup, recent work moves the index into accelerator memory, but the full-vocabulary projection remains unchanged~\citep{su2026static,anandakumar2026flashtrie}.
Some systems further extend these constraints to individual requests, but cannot guarantee that every result satisfies them~\citep{fang2026grace,zhang2025gpr,zhang2026univa}.
All of these methods score first, using the constraint only to discard invalid tokens after the projection has computed the logits of every token rather than to eliminate the redundant computation itself.

\section{Conclusion}
\label{sec:conclusion}

We present \sys, a filtering-first constrained decoding system for generative recommendation.
Instead of scoring every token and then discarding invalid continuations, \sys filters first and scores only valid continuations, with negligible impact on recommendation quality.
It combines a step-sliced filtering index that keeps the index compact in GPU memory, an execution engine whose shared-slot execution and hybrid projection turn reduced scoring into lower latency, and layout-guided streaming construction that rebuilds the index within the update window while limiting interference with online serving.
Compared with representative constrained decoding baselines, \sys reduces the combined latency of filtering, scoring, and candidate selection by up to $9.9\times$, reduces GPU index memory use by up to $74.5\%$, and rebuilds its index $278\times$ faster.

\newpage

\FloatBarrier
\bibliographystyle{ACM-Reference-Format}
\bibliography{references}
\end{document}